\documentstyle[11pt,epsf]{article} 
\begin{document}
\newcommand{\beq}{\begin{equation}}
\newcommand{\eeq}{\end{equation}}
\newcommand{\beqa}{\begin{eqnarray}}
\newcommand{\eeqa}{\end{eqnarray}}
\newcommand{\bea}{\begin{eqnarray}}
\newcommand{\eea}{\end{eqnarray}}
\newcommand{\dfrac}{\displaystyle \frac}
\renewcommand{\thefootnote}{\#\arabic{footnote}}
\newcommand{\ve}{\varepsilon}
\newcommand{\krig}[1]{\stackrel{\circ}{#1}}
\newcommand{\barr}[1]{\not\mathrel #1}
\newcommand{\vs}{\vspace{-0.25cm}}
\newcommand{\no}{\nonumber}

\begin{titlepage}
 
\hfill {\small FZJ-IKP(TH)-1998-29}

\vspace{2.0cm}

\begin{center}

{\Large  \bf {
On the size of isospin violation in \\[0.2em]
low--energy pion--nucleon scattering \footnote{Work supported
    in part by Deutsche Forschungsgemeinschaft under contract 
    no. SCHU-439/10-1.}}}

\vspace{1.2cm}
                              
{\large 
Nadia Fettes$^{\ddag}$\footnote{email: N.Fettes@fz-juelich.de},
Ulf-G. Mei\ss ner$^{\ddag}$\footnote{email: Ulf-G.Meissner@fz-juelich.de},
Sven Steininger$^{\ddag,\dag}$\footnote{email: S.Steininger@fz-juelich.de}
}

\vspace{1.0cm}

{\em 

$^{\ddag}$Forschungszentrum J\"ulich, Institut f\"ur Kernphysik (Theorie)\\
D--52425 J\"ulich, Germany\\

\vspace{0.4cm}
$^{\dag}$Universit\"at Bonn, Institut f\"ur Thoeretische Kernphysik\\ 
Nussallee 14-16, D--53115 Bonn, Germany\\
}

\end{center}

\vspace{2.8cm}

\begin{abstract}
\noindent We present an analysis of isospin--breaking effects in  threshold
pion--nucleon scattering  due to
the light quark mass difference and the dominant virtual photon
effects. We discuss the deviation from various relations,
which are exact in the isospin limit.  The size of the
isospin--violating effects in the relations involving the isovector
$\pi N$ amplitudes is typically of the order of one percent.
We also find a new remarkably large
effect $(\sim 40\%)$ in an  isoscalar triangle
relation connecting the charged and neutral pion scattering off 
protons.

\end{abstract}


\vspace{2cm}


\vfill

\end{titlepage}

\noindent {\bf 1.} Pion--nucleon ($\pi N$) scattering is one of the prime reactions
to test our understanding of the spontaneous and explicit chiral symmetry
breaking QCD is supposed to undergo. During the last years, there has
been  considerable interest in using $\pi N$ scattering
data to extract information about the violation of isospin 
symmetry of the strong interactions~\cite{weincd,bira}, 
some analyses indicating effects as large as 7\%~\cite{gibbs,matsi}. 
In both these analyses, the source of this rather large effect remains
mysterious. Microscopically,
there are two competing sources of isospin violation, which are generally of the
same size, namely the strong effect due to the light quark mass difference
$m_d - m_u \simeq m_u$ and the electromagnetic one caused by virtual
photons. For neutral
pion scattering off nucleons, these effects can be dramatically enhanced due
to the smallness of the isoscalar pion--nucleon amplitude~\cite{wein,ms}.
This spectacular effect in the difference of the $\pi^0 p$ and $\pi^0 n$ scattering 
lengths is, however, at present not amenable to a direct experimental verification.
It is therefore mandatory to include also the channels with charged pions in any
analysis of isospin violation. To do this in a consistent fashion, one has to
develop an effective field theory (EFT) of pions, nucleons and virtual photons. The
corresponding effective Lagrangian was developed in refs.\cite{mms,ms} extending
the standard $\pi N$ EFT (for a review, see ref.\cite{bkmrev}). The pertinent
power counting of the EFT is based on the observation that besides the
pion mass and momenta, the electric charge $e$
should be counted as an additional small parameter, given the fact  
that $e^2/4\pi \simeq M_\pi^2/(4\pi F_\pi)^2 \simeq 1/100$ (with $M_\pi$ and $F_\pi$
the pion mass and decay constant, respectively). From here on, we
collectively denote these small parameters by $q$. Similar information
can also be obtained from precise data on pion photoproduction, as detailed
in ref.\cite{aron} (for an overview, see the talks~\cite{ulfosaka,ulfaus}). The aim of
this paper is to give a first systematic study of the expected size of isospin
violation in  the $\pi N$ amplitude at threshold 
based on a set of relations, which are fulfilled in the limit of exact isospin.
We stress again that in the framework we are using, a consistent
separation of the electromagnetic and the strong effects is possible
and to our knowledge this has not been achieved before. Only when a
mapping of the method developed here on the commonly used procedures of
separating electromagnetic and hadronic mass effects (such as the NORDITA
method~\cite{NO}) has been performed, a sensible comparison with the numbers quoted 
in the literature will be possible.

\medskip

\noindent {\bf 2.} Consider now the process $\pi^a (q_a) + N(p_1) \to 
\pi^b (q_b) + N(p_2)$, where $\pi^a$ denotes a pion of (cartesian)
isospin $a$ and $N$ the nucleon. In the centre--of--mass system (cms),
the four--momentum of the
incoming nucleon is $m \cdot v + p_1 = (E_1 = m + v\cdot p_1, -\vec{q_a})$,
the one of the
outgoing nucleon is $m \cdot v + p_2 = (E_2 = m + v\cdot p_2,
-\vec{q_b})$, where $m$ denotes the nucleon mass in the chiral limit.
Similarly,
the incoming pion has $ q_a = (\omega_a, \vec{q_a}) $ and
the outgoing pion  $ q_b = (\omega_b, \vec{q_b}) $.
The analysis of isospin violation in $\pi N$ scattering
proceeds essentially in three steps. First, one ignores all
isospin breaking effects, i.e. one sets $e= 0$ and $m_u = m_d$.  Only if within
this approximation one is able to describe the low $\pi N$ partial waves in the
threshold region as given by various partial wave analyses, one can be confident
to have a sufficiently accurate starting point.\footnote{Note that in
  the available partial--wave analyses, electromagnetic and some
  hadronic mass effects are
  generally removed by some methods like e.g. the one from NORDITA~\cite{NO}.}
 That this is indeed the case was
demonstrated in refs.\cite{bkmlec,moj,fms}. Ref.\cite{fms}
 also contains a detailed discussion
of the kinematics pertinent to the case considered here. In the second step,
one should include the leading isospin breaking terms encoded in the
pion and nucleon mass differences. The corresponding terms in the effective
Lagrangian read~\cite{mms,ms} (also shown are the terms responsible for the explicit
chiral symmetry breaking)
\beqa\label{eq:L}
{\cal L}_{\pi\pi} &=& \partial_\mu \pi^a \partial^\mu \pi^a + B_0(m_u+m_d)\,\pi^2
+2Ce^2 \,\pi^+ \pi^- + \ldots~,\no \\
{\cal L}_{\pi N} &=& \bar{N} \, ( i\partial_0 + c_1 \, M_{0}^2 (1+\pi^2/F^2)
+ B_0 c_5 (m_u-m_d)\,
\tau^3 (1+\pi^2/F^2) + f_2 e^2 \, \tau^3 + \ldots ) \, N,
\eeqa
where $B_0$ is related to the scalar quark condensate, $B_0=|\langle 0
|\bar{q}q|0\rangle | / F_\pi^2$, and $F$ is the pion decay constant
in the chiral limit, $F_\pi = F[1+{\cal O}(q^2)]$, 
The pion mass difference $M^2_{\pi^+} -
M_{\pi^0}^2$ is entirely determined by the low--energy constant (LEC) 
$C$~\cite{EGPR},\footnote{Note that we work in the $\sigma$--model gauge so that
  the term $C\langle QUQU^\dagger\rangle$ only contributes to
  two--point functions.} 
while to third order in small momenta  the strong (electromagnetic)
proton--neutron mass splitting is given by the LECs $c_5 \, (f_2)$~\cite{ms}. 
Note that the operator $\sim c_5$ does not only
contribute to the strong $np$ mass splitting but also has a 
contribution $\sim \bar NN \pi\pi$ to the two--pion vertex
which will be of relevance later. 
In terms of the operators defined in
eq.(\ref{eq:L}), retaining only the
terms leading to the strong and electromagnetic (em) hadron mass
splitting  is achieved by setting
\beq
Ce^2 \neq 0\, ,\,f_2 e^2 \neq 0\,  , \, m_u - m_d \neq 0\, , \quad
{\rm but }\quad e^2 = 0~.
\eeq 
This is the approximation which we will consider here. In fact, in neutral
pion photoproduction off nucleons, to third order in small momenta, this
approximation leads to the only isospin breaking effect, which reveals itself
in the large cusp effect at the secondary threshold (i.e. at the $\pi^+ n$
threshold in the case of $\gamma p \to \pi^0 p$). In the third step, which
goes beyond the scope of this paper, one has to account for all virtual
photon effects, in particular soft photon emission from charged particle legs
and the Coulomb poles due to the ladder exchange of (hard) virtual photons between
charged external particles. In that case, the notion of partial waves
becomes doubtful and one better compares directly to the available
cross section and polarization data.
We believe, however, that the essential effects
of isospin violation are captured in the calculation presented here. 

\medskip

\noindent {\bf 3.} In the presence of isospin violation, i.e. isovector
symmetry breaking terms such as 
$(m_u -m_d)(\bar{u}u-\bar{d}d)$, one has to generalize the
standard form of the $\pi N$ scattering amplitude to 
\beq 
T^{ab} (\omega , t) =  
\delta^{ab} T^+_{ab}  (\omega , t)
+ \delta^{ab} \tau^3 T^{3+}_{ab}  (\omega , t)
+ i \epsilon^{bac} \tau^c T^-_{ab} (\omega , t)
+ i\epsilon^{bac} \tau^c \tau^3 T^{3-}_{ab} (\omega , t)  
\quad ,
\label{piNscatIV} 
\eeq 
in terms of {\it two isoscalar} ($T^{+,3+}_{ab}$) and {\it two isovector} amplitudes
($T^{-,3-}_{ab}$). These are functions of two variables, here we choose
the pion energy $\omega$ and the invariant  momentum transfer squared $t$.
More precisely, $\omega$ can be chosen to be either the energy of the 
in--coming or out--going pion, since these are no longer equal
\beq\label{Dom}
\Delta \omega  = \omega_b - \omega_a
=  \frac{(M_b^2-M_a^2) - (m_2^2-m_1^2)}{2 \sqrt{s}}
=  \frac{(M_b^2-M_a^2) - (m_2^2-m_1^2)}{2 m_1} 
\left[1 - \frac{w_a}{m_1} + {\cal O}(q^2) \right]~,
\eeq
with $M_{a,b}$ $(m_{1,2})$ the mass of the in--coming, out--going pion
(nucleon) and $\sqrt{s}$
the total cms energy. It is important to note that while the pion
energies $\omega_{a,b}$ are of order $q$, their difference only starts out
at second order in the chiral expansion. This has important
consequences as will be discussed later.
The $T$--amplitudes split, of course, into a spin non--flip
and a spin--flip term, denoted by $g$ and $h$, respectively (for more precise
definitions, see e.g. ref.\cite{fms}). At threshold, only the spin non--flip 
amplitudes can contribute and eq.(\ref{piNscatIV}) simplifies to
\beq\label{Tthr}
T^{ab}_{\pi N, {\rm thr}}= {\cal N}_1 {\cal N}_2 \, \biggl\{
\delta^{ab}\, g_{ab}^+ + \delta^{ab}\,\tau^3 \,g_{ab}^{3+}
+  i\, \epsilon^{bac} \, \tau^c \, g_{ab}^- 
+  i\, \epsilon^{bac} \, \tau^c\, \tau^3 \, g_{ab}^{3-}~\biggr\},
\eeq
with ${\cal N}_i =\sqrt{(E_i + m_i)/2m_i}\,\, (i=1,2)$ the standard spinor
normalization (its relevance is discussed in detail in ref.\cite{fmsJ}).
In what follows, we do not consider these
normalization factors since they are related to the external kinematics.
Consequently, the isospin violating effects are entirely confined to happen
within the given Feynman graphs we consider. 
The $g_{ab}^{\pm,3\pm}$ are, of course, proportional to the corresponding
S--wave scattering lengths. Note that in the presence of isospin violation,
these amplitudes can become complex even at threshold (since the mass of the incoming 
two--particle system is no longer equal to the mass of the out--going one).
It is also important to realize that the $g_{ab}^{\pm}$ are exclusively
sensitive to the neutral to charged pion mass difference (i.e. the LEC $C$)
whereas the $g_{ab}^{3\pm}$ are given by the operators $\sim f_2, \sim c_5$,
i.e the strong {\it and} em proton--neutron mass difference.

\medskip

\noindent {\bf 4.} Isospin violation is best characterized in terms of
quantities which are exactly zero in the isospin limit of equal quark
masses and vanishing em coupling. With the three pion ($\pi^\pm,
\pi^0)$ and two nucleon ($p,n$) fields, we have ten reaction
channels, which in the case of isospin symmetry are entirely described
in terms of two amplitudes. One thus
can write down eight isospin relations (see also ref.\cite{kg} for
a general analysis)
\bea 
R_1 & = & 
2 \, \frac{T_{\pi^+ p \to \pi^+ p} + T_{\pi^- p \to \pi^- p} 
- 2 \, T_{\pi^0 p \to \pi^0 p}} 
          {T_{\pi^+ p \to \pi^+ p} + T_{\pi^- p \to \pi^- p} + 2 \,
            T_{\pi^0 p \to \pi^0 p}} \no \\ &=&  
2 \, \frac{g^+_{11}+g^+_{22}-2 \, g^+_{33}+g^{3+}_{11}+g^{3+}_{22}-2 \, g^{3+}_{33}} 
          {g^+_{11}+g^+_{22}+2 \, g^+_{33}+g^{3+}_{11}+g^{3+}_{22}+2 \, g^{3+}_{33}}~, 
\label{IRel1} \\
R_2 & = & 
2 \, \frac{T_{\pi^+ p \to \pi^+ p} - T_{\pi^- p \to \pi^- p} 
- \sqrt{2} \, T_{\pi^- p \to \pi^0 n}} 
          {T_{\pi^+ p \to \pi^+ p} - T_{\pi^- p \to \pi^- p} 
+ \sqrt{2} \, T_{\pi^- p \to \pi^0 n}} 
\no \\
& = & 
2 \,
\frac{g^-_{12}+g^-_{21}-g^-_{13}-g^-_{23}+g^{3-}_{12}+g^{3-}_{21}-
g^{3-}_{13}-g^-_{23}} 
          {g^-_{12}+g^-_{21}+g^-_{13}+g^-_{23}+g^{3-}_{12}+g^{3-}_{21}+
g^{3-}_{13}+g^{3-}_{23}}~, 
\label{IRel2} \\  
R_3 & = & 
2 \, \frac{T_{\pi^0 p \to \pi^+ n} - T_{\pi^- p \to \pi^0 n}} 
          {T_{\pi^0 p \to \pi^+ n} + T_{\pi^- p \to \pi^0 n}} 
\no \\
& = & 
2 \, \frac{g^-_{31}+g^-_{32}-g^-_{13}-g^-_{23}+g^{3-}_{31}+
g^{3-}_{32}-g^{3-}_{13}-g^{3-}_{23}} 
          {g^-_{31}+g^-_{32}+g^-_{13}+g^-_{23}+g^{3-}_{31}+
g^{3-}_{32}+g^{3-}_{13}+g^{3-}_{23}}~, 
\label{IRel3} \\  
R_4 & = & 
2 \, \frac{T_{\pi^+ p \to \pi^+ p} - T_{\pi^- n \to \pi^- n}} 
          {T_{\pi^+ p \to \pi^+ p} + T_{\pi^- n \to \pi^- n}} 
=  
2 \, \frac{g^{3+}_{11}+g^{3+}_{22}-g^{3-}_{12}-g^{3-}_{21}} 
          {g^+_{11}+g^+_{22}-g^-_{12}-g^-_{21}}~, 
\label{IRel4} \\
R_5 & = & 
2 \, \frac{T_{\pi^- p \to \pi^- p} - T_{\pi^+ n \to \pi^+ n}} 
          {T_{\pi^- p \to \pi^- p} + T_{\pi^+ n \to \pi^+ n}} 
= 
2 \, \frac{g^{3+}_{11}+g^{3+}_{22}+g^{3-}_{12}+g^{3-}_{21}} 
          {g^+_{11}+g^+_{22}+g^-_{12}+g^-_{21}}~, 
\label{IRel5} \\
R_6 & = & 
2 \, \frac{T_{\pi^0 p \to \pi^0 p} - T_{\pi^0 n \to \pi^0 n}} 
          {T_{\pi^0 p \to \pi^0 p} + T_{\pi^0 n \to \pi^0 n}} 
 =  
2 \, \frac{g^{3+}_{33}} 
          {g^+_{33}}~, 
\label{IRel6} \\
R_7 & = & 
2 \, \frac{T_{\pi^- p \to \pi^0 n} - T_{\pi^+ n \to \pi^0 p}} 
          {T_{\pi^- p \to \pi^0 n} + T_{\pi^+ n \to \pi^0 p}} 
 =  R_3~, 
\label{IRel7} \\
R_8 & = & 
2 \, \frac{T_{\pi^0 p \to \pi^+ n} - T_{\pi^0 n \to \pi^- p}} 
          {T_{\pi^0 p \to \pi^+ n} + T_{\pi^0 n \to \pi^- p}} 
 = -R_3~.
\label{IRel8} 
\eea 
The first two, the
so--called triangle relations, are based on the observation that in the
isospin conserving case, the elastic scattering channels involving charged
pions are trivially linked to the corresponding neutral pion elastic scattering or
the corresponding charge--exchange amplitude.
To be precise, these ratios are to be formed with the real parts of
the corresponding amplitudes evaluated at the pertinent threshold,
symbolically $T_{\pi^a N\to \pi^b N}$ should read ${\rm Re}~T_{\pi^a N\to \pi^b
  N}^{\rm thr}$. The imaginary parts of some of the amplitudes
will be discussed later. 
Of particular interest is the second ratio, which is often 
referred to as {\it the triangle relation}. Only in this case all three
channels have been measured (for pion kinetic energies as low as 30~MeV in the
cm system) and the 7\% strong isospin violation reported in refs.\cite{gibbs,matsi}
refers to this ratio. We stress again
that it is difficult to compare this number with the one obtained in
our calculation since a very different method of separating the em
effects is used. The ratio $R_6$ parametrizes
the large isospin violation effect for $\pi^0$ scattering off nucleons
first found by Weinberg~\cite{wein} and
sharpened in ref.\cite{ms}, $R_6 \simeq 25\%$. Note that in $R_1$ the
isovector terms drop out completely and one thus expects also a large
isospin violation in this ratio (since the isoscalar parts are strongly
suppressed and are of the same size as  the symmetry breaking terms).
To our knowledge, this is the first time  that this particular ratio
has been called attention to. From an experimental point of view, it
has the advantage of avoiding the almost unmeasurable $n\pi^0$
amplitude appearing in $R_6$. However, both $R_1$ and $R_6$ are sensitive to the precise
values of the combination of LECs $c_2+c_3-2c_1$ 
since the strong contribution to the
isoscalar scattering length is not even known in sign at present.
The predictions for the other ratios are more stable since they
involve the larger (and better determined) isovector quantities.
Note that the relations $R_7 = R_3 = -R_8$ follow
from time reversal invariance. In what follows, we will
calculate the six ratios $R_i$ to leading one loop accuracy, i.e. to third order
in small momenta. For that, we have to consider tree graphs, some with
fixed coefficients and some with LECs, and the leading one loop graphs
involving lowest order couplings only.

\medskip

\noindent{\bf 5.} The pertinent Born graphs calculated to first, second and
third order are depicted in fig.2 of ref.\cite{fms}. The ones contributing
here are 1b, 2a, 2b, 2d, 3a-3f and the additional tree graphs $\sim f_2, \sim
c_5$ are shown in fig.1. Before giving the results for the
sum of all Born graphs (tree graphs with or without LECs), some important
remarks concerning the chiral power counting are in order. Although  
the so--called Weinberg--Tomozawa $\bar{N}N\pi\pi$ contact graph gives
a first order contribution to $g_{ab}^-$, in the ratios $R_i$ its
effect is always proportional to  $\Delta \omega$, which is of second order,
cf. eq.(\ref{Dom}).
Consequently, isospin violation only starts at second order in the chiral
expansion. Furthermore, in some tree graphs with intermediate nucleon lines
the pion energy difference enters since $v\cdot p_1 - v\cdot p_2 
= \omega_b - \omega_a$,
which is thus also of order $q^2$. We have only accounted for this difference whenever
it was necessary and consequently neglected it when it would only lead to
a fourth (or higher) order contribution. This applies in particular also to the loop 
graphs given below. Therefore, the final results depend on the choice of taking either
the energy of the in--coming or the out--going pion as 
reference energy. This difference is,
however, beyond the accuracy we are working and thus gives us the possibility
to estimate some higher order effects. After mass and coupling constant
renormalization, the sum of all Born graphs gives the following
contributions (for notations, see ref.\cite{fms})
\beqa
F_\pi^2 \, g^{+}_{ab} 
&=& -4c_1 M_0^2 + 2(c_2 - \frac{g_A^2}{8m}) \omega^2
+ 2c_3 \omega^2~,\\
F_\pi^2 \, g^{3+}_{ab} 
&=& -2B_0(m_u -m_d) c_5 \delta^{a3} + \frac{1}{2} e^2
f_2 F_\pi^2 (\delta^{a3} -1)~,\\
F_\pi^2 \, g^{-}_{ab} &=& \frac{\omega_a + \omega_b}{4}
+\frac{|\vec{q}_a|^2+|\vec{q}_b|^2}{8m}(1-2g_A^2)
+4\omega \bigl[ \omega^2(d_1+d_2+d_3)+2M_0^2 d_5 + M_0^2 \tilde{d}_{28}
\biggr] \no \\
&+& \frac{\omega(1-4g_A^2)}{16m^2}\biggl( |\vec{q}_a|^2+|\vec{q}_b|^2 \biggr)
+\frac{g_A^2 \omega^3}{8m^2} - \frac{\omega}{2m}(c_4 + \frac{1}{8m})
\biggl( |\vec{q}_a|^2+|\vec{q}_b|^2 \biggr)~, \\
F_\pi^2 \, g^{3-}_{ab} &=& 
\biggl\{-B_0(m_u -m_d) c_5  + \frac{1}{4} e^2
f_2 F_\pi^2 \biggr\} (\delta^{a3} - \delta^{b3}) \no \\
&+& \frac{g_A^2}{2m\omega} ( |\vec{q}_a|^2+|\vec{q}_b|^2)
\biggl\{2B_0(m_u -m_d) c_5  + \frac{1}{2} e^2 f_2 F_\pi^2 \biggr\} 
(\delta^{a3} - \delta^{b3})~.
\eeqa
At third order, we have to consider the set of one loop graphs shown
in fig.2. These give 
\beqa
F_\pi^4 \, g^{+}_{ab} 
&=& -\frac{\omega^2}{4}
\left(J_0^c(\omega)+J_0^c(-\omega)\right)
(\delta^{ac}-\delta^{cc})
-\frac{g_A^2}{4} \delta^{cc} 
M_0^2 (J_0^c(0)-M_c^2 \gamma_0^{cc}(0)) \no \\
&-&
\frac{g_A^2}{2} 
\left( 
M_a^2 (2M_a^2-M_0^2) \gamma_0^{aa}(0)
+ (M_0^2-3M_a^2) J_0^a(0))
\right) 
- \frac{g_A^2}{2}
M_a^2 J_0^a(0)~, \\
F_\pi^4 \, g^{-}_{ab} &=&
\frac{\omega}{16}
\left(
4\omega
\left(J_0^c(\omega)-J_0^c(-\omega)\right)
+ 6\Delta_\pi^c
\right) \no \\
&+& \frac{\omega}{12 t}
\Bigg(
(t^2-2t(M_a^2+M_b^2)+(M_a^2-M_b^2)^2) I_0^{ab} (t)
\no \\
& & \left.
+ t (\Delta_\pi^a + \Delta_\pi^b)
+ (M_a^2-M_b^2) (\Delta_\pi^a - \Delta_\pi^b)
+ \frac{t^2-3t(M_a^2+M_b^2)}{24\pi^2}
\right) \no \\
&+& \frac{g_A^2}{48}
\Bigg\{
4\omega [5t-4(M_a^2+M_b^2)-\frac{(M_a^2-M_b^2)^2}{t}] I_0^{ab}(t)
+20\omega(\Delta_\pi^a+\Delta_\pi^b)
\no \\ & &
-\frac{4\omega}{t}(M_a^2-M_b^2)(\Delta_\pi^a-\Delta_\pi^b)
-\frac{\omega}{6\pi^2}(t-3(M_a^2+M_b^2))
\Bigg\} \no \\
&+& \frac{\omega}{8}
\left(
\delta^{dd} \Delta_\pi^d + \Delta_\pi^a + \Delta_\pi^b
\right) - \frac{g_A^2}{8}
\omega\left(
M_d^2 J_0^{d\,\prime}(0)-\Delta_\pi^d
\right)
(2\delta^{dc}-\delta^{dd}) \no \\
&+&\frac{\omega_a + \omega_b}{4} \left(
\bigl(\frac{g_A}{F}\bigr)^2 \frac{1}{4} (M_c^2 J_0^{' c}(0) - \Delta_\pi^c)
+ 8 M_0^2 d_{28}(\lambda)
-\frac{1}{F_\pi^2}\delta^{c c} \Delta_\pi^c \right)~,
\eeqa
where one has to generalize the standard loop functions
$I_0, \Delta_\pi, \gamma_0, J_0$ (see the
appendices in refs.\cite{bkmrev,GL85}) to the unequal mass case as
indicated by the superscripts $'a,b,c\, '$ referring to the pion isospin
indices. The loop contributions to $g_{ab}^-$ are divergent. These
divergences are cancelled by the appropriate dimension three operators
from ${\cal L}_{\pi N}^{(3),\rm em}$ as constructed in ref.\cite{ms}.
Note that in the approximation we are using, the finite parts of these
terms are set to zero. We remark that the imaginary parts generated by
the loops are very small for the threshold kinematics considered here.
This is made more precise in the following.

\medskip

\noindent {\bf 6.} We are now in the position to analyze the ratios $R_i$
as defined in eqs.(\ref{IRel1}-\ref{IRel8}). We use the standard masses
as given in the PDG tables for $m_{p,n}, M_{\pi^{0,\pm}}$~\cite{PDG}. For the mass
parameter $m$
we can use $m = (m_p + m_n)/2$ to the order we are working and we identify
the leading term in the quark mass expansion of the pion mass with the
neutral pion mass, $M_0 = M_{\pi^0}$. 
We also use  $F_\pi = 92.4\,$MeV\footnote{In principle, we would have
to differentiate between $F_{\pi^{+}}$ and $F_{\pi^{0}}$. This difference is
of second order and would therefore show up as a third order contribution
due to the Weinberg--Tomozawa term. At present, the empirical determinations
do not allow to differentiate between these two values and we thus work with
one value given by the charged decay constant.} and $g_A = 1.26$. 
The LECs $c_{1,2,3,4}$ and $d_i$ are taken from fit~1,2,3\footnote{Fit
1,2 and 3 refers to the
Karlsruhe~\cite{koch}, the Matsinos~\cite{mats} and the 
VPI~\cite{SAID} partial wave analysis, in order.} of
ref.\cite{fms} (we refer to that paper as ``FMS''), $c_5$ from 
ref.\cite{bkmlec} and $f_2$ from ref.\cite{ms}.  

In table~\ref{tab:Ri}, we give the results for the ratios $R_i$ that
are not entirely given by isoscalar quantities. These numbers should
be more stable than the ones for the isoscalar ratios $R_{1,6}$. This
is indeed the case, there are no large variations between the three
parameter sets given in FMS. For the pion energy $\omega$, we have used the arithmetic
mean of the in--coming and out--going energies (this is also the most
natural choice since it preserves the time reversal invariance between
$\pi^+ n \leftrightarrow \pi^0 p$ and $\pi^- p \leftrightarrow \pi^0 n$).
For parameter set~1, we have also varied the pion reference energy and used
either $\omega_a$ or $\omega_b$. The $R_{1,4,5,6}$ are insensitive
to this choice, whereas $R_{2,3}$ can vary between approximately zero
and 1.5\%. This points towards the necessity of a fourth order
calculation. 
\renewcommand{\arraystretch}{1.2}
\begin{table}[htb]
\begin{center}
\begin{tabular}{|c|c|c|c|c|}
    \hline
   & $R_2$ [\%]  &  $R_3$ [\%] &  $R_4$  [\%] & $R_5$ [\%]  \\
    \hline\hline
Fit 1 & $0.9$ & $-0.5$ & $-0.7$ & $1.1$ \\
Fit 2 & $1.1$ & $-0.6$ & $-0.9$ & $1.1$ \\
Fit 3 & $0.9$ & $-0.5$ & $-0.8$ & $1.0$ \\
   \hline\hline
  \end{tabular}
  \caption{Values of the ratios $R_i$ ($i=2,3,4,5)$  
           for the various parameter sets as given by the
           fits of FMS.\label{tab:Ri}}
\end{center}\end{table}

We now turn to the two isoscalar ratios. First, we consider $R_6$,
which was first discussed by Weinberg~\cite{wein}. For fit~1, we find $R_6
= 19$\%, which is somewhat smaller than the 25\% reported in
ref.\cite{ms}. Note, however, that the value for $a^+$  based on
the KA85 phase shifts is larger (in magnitude) than the one used
in ref.\cite{ms} (based on the LECs as determined in ref.\cite{bkmlec})
 and thus the isospin violating effect
is indeed expected to be smaller. On the other hand, for the 
parameters of fits~2 and 3,
$R_6$ gets much larger, but not because the isospin--violating
function $g^{3+}_{33}$ changes (it is indeed stable up to fourth order
as argued in ref.\cite{ms}), but rather the isospin--conserving
function $g^{+}_{33}$ varies considerably. Thus, to sharpen the
prediction for $R_6$, one has to go to next order in the
isospin--conserving case. Interestingly, for the same parameter set
(fit~1 of FMS), the prediction for $R_1$ is even larger,
\beq\label{R1v}
R_1 = 36.7\%~,
\eeq
which is again a huge isospin violating effect in an isoscalar
quantity. The same remarks as made for $R_1$ apply here. However,
we stress again that this novel ratio could be accessible
experimentally if the proposal of Bernstein~\cite{aron} to extract the
$\pi^0 p$ scattering by precise pion photoproduction experiments could
be carried out. Again, we remark that the value for $R_1$ given in eq.(\ref{R1v})
should be considered as a lower limit since for the other parameter
sets the isospin--conserving isoscalar amplitude is smaller (in
magnitude) which enhances the very stable isospin--breaking difference
in the ratio. It is thus mandatory from theory and experiment to get
a more precise value for the isoscalar amplitude. 

It is most interesting to separate  the hadronic isospin violation
encoded in the operator $\sim c_5 (m_u-m_d)$ from the virtual photon
effects. One could set 
$c_5 = 0$, i.e. all strong isovector terms would vanish. This is not quite
what one wants since then the proton is heavier than the neutron. 
In that case, the masses of the external particles would also change
and a meaningful comparison becomes difficult. We
can, however, keep $c_5 \neq 0$ for the nucleon mass insertions and
set it to zero in the $\bar N N \pi \pi$ terms $\sim c_5$, cf. eq.(\ref{eq:L}),
which we denote by $c_5 (\pi\pi )=0$ in table~2. This splitting of the 
strong isospin violating terms is similar to what is called ``static'' and
``dynamical'' isospin breaking in ref.\cite{aron}. We see that while
the em isospin breaking is generally dominant (with the exception of
$R_6$), there is also some sizeable strong isospin breaking.
\renewcommand{\arraystretch}{1.2}
\begin{table}[htb]
\begin{center}
\begin{tabular}{|l|c|c|c|c|c|c|}
    \hline
   & $R_1$ & $R_2$  &  $R_3$ &  $R_4$  & $R_5$ & $R_6$  \\
    \hline\hline
$c_5 \neq 0$     & $36.7$ & $0.9$ & $-0.5$ & $-0.7$ & $1.1$ & $19.3$ \\
$c_5 (\pi\pi ) =0$ & $45.4$ & $1.6$ & $ 0.8$ & $-0.7$ & $1.1$ & $ 0  $ \\
   \hline\hline
  \end{tabular}
  \caption{Values of the ratios $R_i$ ($i=1,2,3,4,5,6)$  
           for the parameters of fit~1 from FMS including the
           full contribution from the strong isospin breaking ($c_5
           \neq 0$) and
           with the strong isospin violation only contributing to the
           proton--neutron mass difference ($c_5 (\pi\pi ) =0$).
           \label{tab:Ric5}}
\end{center}\end{table}

So far, we have discussed the ratios $R_i$ from the real parts of
the complex--valued 
$\pi N$ amplitudes evaluated at the pertinent
threshold kinematics. It is also instructive to give the 
corresponding scattering lengths. We define these as the amplitude
at threshold (i.e. in some cases we have complex numbers) including
the normalization factors, symbolically $a = \sqrt{m_1 m_2}T_{\rm thr}
/ (4\pi \sqrt{s})$. For
scattering pions off protons, we get the numbers given in table~3. 
The imaginary part in $a(\pi^- p \to \pi^- p)$ is due to the
intermediate $\pi^0 n$ state, whereas for $a(\pi^0 p \to \pi^+ n)$
the energy of the initial two--body system is smaller than the one of
the final two--body state. In both cases, these imaginary parts are
fairly small. As a check on the numerics, we find that
$a(\pi^+ n \to \pi^0 p) = a(\pi^0 p \to \pi^+ n)$ and $a(\pi^- p \to
\pi^0 n) = a(\pi^0 n \to \pi^- p)$ as demanded by time reversal invariance.
For comparison, the scattering lengths in the isospin limit (using the charged pion
and the proton mass) can be obtained from table~2 of FMS by use of  the
relations $a(\pi^\pm p \to \pi^\pm p) = a_{0+}^+ \mp a_{0+}^- $,
$a(\pi^0 p \to \pi^0 p) =  a_{0+}^+$ and $a(\pi^- p \to \pi^0 n) =
a(\pi^0 p \to \pi^+ n) = -\sqrt{2} a_{0+}^-$.
\renewcommand{\arraystretch}{1.2}
\begin{table}[htb]
\begin{center}
\begin{tabular}{|c|c|c|c|c|c|}
    \hline
   & $a(\pi^+ p \to \pi^+ p)$  &  $a(\pi^- p \to \pi^- p)$ &  
   $a(\pi^0 p \to \pi^0 p)$ & $a(\pi^- p \to \pi^0 n)$ & $a(\pi^0 p \to \pi^+ n) $\\
    \hline\hline
Fit 1 & $-108.7$& $70.2 + i\,3.65$& $-13.4$& $-125.5$& $-124.7 -i\,0.63$ \\
Fit 2 & $-83.8$ & $71.3 + i\,3.65$& $-0.1$ & $-108.6$& $-107.9 -i\,0.63$ \\
Fit 3 & $-94.9$ & $77.7 + i\,3.65$ & $-1.8$& $-121.1$& $-120.2 -i\,0.63$ \\
   \hline\hline
  \end{tabular}
  \caption{Values of the scattering lengths for pion scattering of
   protons in units of $10^{-3}/M_{\pi^+}$
   for the various parameter sets as given by the
   fits of FMS.\label{tab:a}}
\end{center}\end{table}

\medskip

\noindent {\bf 7.} In this paper, we have considered isospin violation in 
low energy pion--nucleon scattering in the framework of heavy baryon chiral
perturbation theory to third order in small momenta. We have taken into account
all operators related to strong isospin breaking and the electromagnetic ones,
which lead to the pion and nucleon mass differences. Stated differently, the finite
parts of some of the virtual photon operators contributing at this order have been
set to zero. This allows in particular to isolate the contribution of the strong
dimension two isovector operator first considered by Weinberg. We have considered
a set of six ratios $R_i$, which vanish in the limit of isospin conservation.
{}From these, six involve isovector {\it and} isoscalar amplitudes ($R_{2,3,4,5,7,8}$) 
and the two others are purely of isoscalar type $(R_{1,6})$. While in the first
case, isospin violation is typically of the order of one percent, more sizeable
effects are found in $R_6$~\cite{wein,ms} and, as for the first time noted here,
in $R_1$. These results strongly motivate  efforts to measure more precisely
the isoscalar scattering length $a^+$ and try to determine the $\pi^0 p$ scattering
length e.g. from accurate threshold pion photoproduction experiments. We also stress
again that within the framework presented here, a unique and unambiguous separation
of all different isospin violating effects is possible. To access the size of
isospin violation encoded in the presently available pion--nucleon scattering data,
the extension of this scheme to include Coulomb (hard) and soft photons is mandatory.
Once this is done, it will be possible to analyze directly the cross
section data without 
recourse to any model for separating em or hadronic mass effects thus, avoiding any
mismatch by combining different approaches or models. Work along such lines is underway.

\newpage

\noindent {\Large {\bf Figures}}

$\,$

\vskip 1.5cm

\begin{figure}[htb]
   \vspace{0.5cm}
   \epsfxsize=8cm
   \centerline{\epsffile{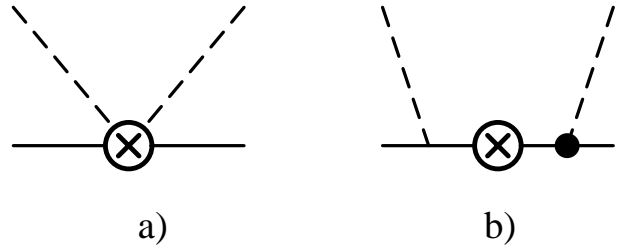}}
   \vspace{0.2cm}
   \centerline{\parbox{15cm}{\caption{\label{fig1}
   Tree graphs contributing to isospin violation in  $\pi N$ 
   scattering. Solid and dashed lines denote nucleons and pions, in
   order. The circle--cross (heavy dot) refers to a dimension two
   insertions $\sim f_2$ or  $\sim c_5$ ($\sim 1/m$). 
}}}
\end{figure}

$\,$

\vskip 1cm

\begin{figure}[htb]
   \vspace{0.5cm}
   \epsfxsize=12cm
   \centerline{\epsffile{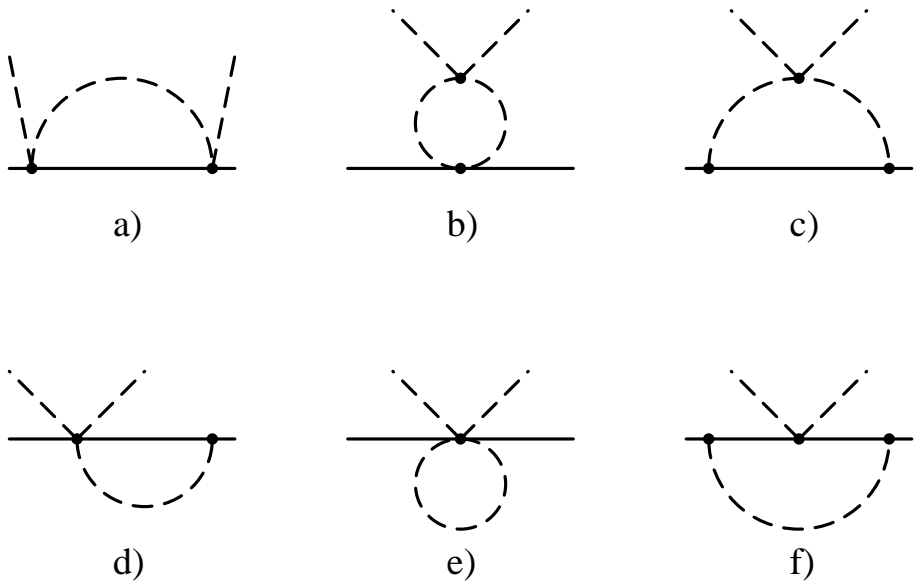}}
   \vspace{0.2cm}
   \centerline{\parbox{15cm}{\caption{\label{fig2}
   One loop graphs contributing to isospin violation in  $\pi N$ 
   scattering. Solid and dashed lines denote nucleons and pions, in
   order. 
}}}
\end{figure}

\end{document}